\documentstyle[sprocl,epsfig]{article}

\def\lesssim{\mathrel{\hbox{\rlap{\hbox{\lower4pt\hbox{$\sim$}}}\hbox{$<$}}}}
\def\beq{\begin{equation}}
\def\eeq{\end{equation}}
\def\beqa{\begin{eqnarray}}
\def\eeqa{\end{eqnarray}}

\begin{document}
\title{Recent Progress in Electroweak Baryogenesis}

\author{James M.\ Cline}
\address{Department of Physics, McGill University\\
Montr\'eal, Qu\'ebec H3A 2T8 Canada\\ E-mail: jcline@physics.mcgill.ca}


\maketitle\abstracts{ The status of electroweak baryogenesis in the
minimal supersymmetric standard model is reviewed.  I discuss the
strength of the phase transition, CP violation and transport at the
bubble walls, and the possibility of a two stage transition involving
charge and color breaking.}

\vskip-2.5in
\rightline{McGill/99-04}
\vskip2.25in

\section{Electroweak Baryogenesis: Review}

The basic ideas underlying electroweak baryogenesis
\cite{CKN,HN} have been
reviewed elsewhere,\cite{reviews} so I will give an abbreviated
description here in order to leave room for some of the interesting
details.  To create the baryon asymmetry of the universe during the
electroweak phase transition (EWPT), it is assumed that Sakharov's
out-of-equilibrium requirement \cite{Sakharov} for baryogenesis is
fulfilled by making the EWPT first order:  bubbles of nonzero Higgs
field VEV nucleate from the symmetric vacuum and fill up the universe.
As these bubbles expand, there is a flux of particles going into the
bubbles, but there is also some probability for the particles to bounce
off the wall back into the symmetric phase, where their masses are
smaller.  If there are new CP violating effects in the bubble walls,
this reflection probability can be different for left-handed quarks and
their antiparticles, for example.  This produces a CP-asymmetric flux
of particles back into the symmetric phase.  Baryon number violating
sphaleron interactions are biased by this CP asymmetry into producing a
baryon asymmetry in front of the wall.  Since the wall is steadily
advancing, whereas particles in the vicinity of the wall are diffusing,
these baryons make their way inside the bubbles, where the sphaleron
interaction rate is much smaller.  As long as sphalerons are
ineffective in the interior, these baryons survive to become the
visible matter of the present universe.  The rest is history, one might
say.

\section{Strength of the Phase Transition}

Not only is new physics needed to get enough CP violation, but the
hypothesis of a strongly first order electroweak phase transition also
requires it.  By strong we mean that the Higgs VEV in the broken phase, 
$v_c$, at the critical temperature, $T_c$, must be sufficiently large,
\beq
\label{crit}
	v_c/T_c > 1.
\eeq
Otherwise, sphaleron interactions continue to be too fast inside the
bubbles and destroy any baryon asymmetry which gets produced.  If the
top quark and Higgs boson were both light, this condition could be
achieved within the SM, but it is now well established not only that
the transition is too weak, but it is smooth, with no bubbles at
all.\cite{lattice}\ \ To get a first order transition, one needs a
negative cubic term in the finite-temperature effective Higgs
potential,
\beq
\label{potential}
V(H) = \frac12(-\mu^2_h + c_h T^2) H^2 - E T H^3 + \frac14 \lambda_h H^4 .
\eeq
At the critical temperature, where $V(H)$ takes the form $(\lambda/4)
H^2 (H-v_c)^2$, it is easy to show that \vspace{-1pt}
\beq
\label{Tc}
	{v_c\over T_c} = {2E\over\lambda}.
\eeq
Thus one needs not only for $E$ to be large, but also 
$\lambda$ must be small,
which implies a small mass for the Higgs boson.
The cubic term in (\ref{potential}) 
arises from the one-loop free energy for bosons,
which at high temperatures goes like 
\beq
  V(m(H)) = \frac{1}{24} m^2 T^2 - \frac{1}{12\pi} m^3 T + O(m^4)
\eeq
per degree of freedom, where $m(H)$ is the Higgs field dependent boson
mass.  In the SM, such cubic contributions are provided by the gauge
bosons, but they are too small to make $v_c/T_c >1$.  However new
particles which couple strongly enough to the Higgs field can increase
the cubic coupling, provided that they are also sufficiently light.

For those who look to supersymmetry for new physics, the natural
candidate for strengthening the transition is the right top squark
(stop), since it has the strong top Yukawa coupling $y^2 |\tilde t_R|^2
|H_2|^2 $ to the second of the two Higgs doublets required by SUSY.  We
focus on $\tilde t_R$ because the left stop tends to make excessive
contributions to the electroweak $\rho$ parameter if it is light. If we
write $H = \sqrt{H_1^2 + H_2^2}$ and $H_2 = \sin\beta H$, the
field-dependent stop mass has the form 
\beq
	m_{\tilde t_R}^2 = m^2_U + c_s T^2 + y'^2 H^2,
\label{stopmass}
\eeq
where $y'$ is less than $y\sin\!\beta$ when mixing between
left and right stops is accounted for.  But to get a cubic term $H^3$
from {\small $m_{\tilde t_R}^3$}, it is necessary that the soft SUSY-breaking
mass $m^2_U$ be negative \cite{CQW1} so as to approximately cancel the
thermal contribution $c_s T^2$.  At zero temperature, the physical mass of the
stop will then be less than $y' v$, which in turn is less than the top
quark mass.  To be more precise, $c_s \cong 0.5$ and $T \cong 90$ GeV,
so $m_{\tilde t_R} \cong 160$ GeV in the no-mixing case, and
arbitrarily smaller (down to the experimental limit of 85 GeV) with
mixing between the left and right stops.

\begin{figure}[h!] 
\centerline{\epsfig{file=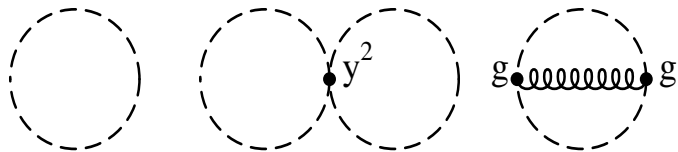,height=0.75in,width=3.5in}}
\caption{Some of the virtual squark diagrams contributing to $V(H)$.}
\label{loopfig}
\end{figure}

It was first pointed out by Espinosa \cite{JRE} that two-loop contributions
to $V(H)$, such as those shown in figure \ref{loopfig}, 
also make a considerable improvement in the strength of the transition.
There have now been a number of studies of the phase transition using
the two-loop effective potential, \cite{BJLS}$^-$\cite{ML} which obtain
similar results for the range of Higgs and stop masses that are 
compatible with having a strong enough phase transition,
\begin{eqnarray}
\label{eq2}
	85 {\rm\ GeV} &<& m_h < 107-116 {\rm\ GeV}\, ;\\
	120 {\rm\ GeV} &<& m_{\tilde t_R} < 172 {\rm\ GeV}\, .\nonumber
\end{eqnarray}
The range of values for the upper limit of the Higgs mass is due to our
freedom to vary the mass ($m_Q$) of the left stop,\cite{CM} which
radiatively corrects $m_h$.  We somewhat arbitrarily considered upper
limits of $m_Q < 1-2$ TeV to obtain this range.  Although even larger
values of $m_Q$ are conceivable, it is difficult to explain why there
should be such a vast hierarchy between the left and right stop
masses.

The fact that two-loop effects are so important, and that the usual
loop expansion does not generally converge well at high temperature,
are cause for concern as to whether reliable results can be obtained
from the two-loop potential.  Fortunately we have empirical evidence
that the two-loop approximation is good, because lattice gauge theory
simulations of the phase transition have been done \cite{LR} which agree
reasonably well with the analytic results; the lattice gives values of
$v_c/T_c$ about 10\% higher than those of the effective potential, for
the limited range of MSSM parameters where the comparison could be
made.  

\subsection{Studying the EWPT in the MSSM}

In this subsection I will summarize the results of ref.\ [12],
whose goal was to make a broad search of the MSSM parameter space for
those values giving a strong enough phase transition, in the sense
of inequality (\ref{crit}).  The main tasks were the following:
\begin{enumerate}

\item We generalized the two-loop effective potential of ref.\ [9]
to include mixing between the various Higgs bosons and between the
left and right stops and sbottoms.  This allowed us to check the cases
where the CP-odd Higgs boson $A^0$ or the left stop and sbottom were
light, rather than decoupled from the thermal bath.

\item For each set of trial MSSM parameters, we checked the
experimental constraints on the light Higgs and squark masses,
$\tan\beta$, the $\mu$ parameter (which is the $\mu H_1 H_2$ coupling
in the superpotential), the electroweak $\rho$ parameter, and the
prohibition against charge- or color-breaking minima due to
condensation of the stop field.  More will be said about the latter in
section 4.

\item For sets that passed the above cuts, we found the critical
temperature $T_c$, not from the approximation (\ref{Tc}) but by tuning
$T$ to get degenerate minima of the full potential.  We then found the
bubble nucleation temperature, $T_{\rm nuc}< T_c$, where one bubble starts
to appear per Hubble volume and Hubble time; we also computed the latent
heat of the phase transition, which allows one to find the reheating
temperature $T_r$, somewhere between $T_{\rm nuc}$ and $T_c$.

\item To accurately find the sphaleron rate, $\Gamma_{\rm sph}$, we
numerically solved for the energy of the sphaleron configuration in the
full potential at $T_r$.  $\Gamma_{\rm sph}$ is given by $(v_c/T_r)^7
e^{6.9 - E_{\rm sph}/T_r}$.  This was compared to a critical allowed
rate, $\Gamma_{\rm crit}$, roughly equal to the expansion rate of
the universe, rather than using the approximation (\ref{crit}), to
determine if the phase transition was strong enough to avoid baryon
dilution by sphalerons.

\item For each accepted parameter set, we computed the profile of the
bubble wall,\cite{MQS} {\it i.e.,} $H_1(r)$ and $H_2(r)$, which enables
one to find the spatial variation of $\tan\beta(r) = H_2/H_1$.  This
quantity was of interest because the efficiency of electroweak
baryogenesis is predicted by some \cite{CQRVW} to be proportional
$\partial\beta/\partial r$. 

\item These steps were carried out for approximately 10,000 randomly
chosen sets of parameters, where we varied $\tan\beta$, the
soft-SUSY-breaking squark masses and mixing parameters, and the mass of
the CP-odd Higgs boson, $m_{A^0}$.

\end{enumerate}

\begin{figure}[h!] 
\centerline{\epsfig{file=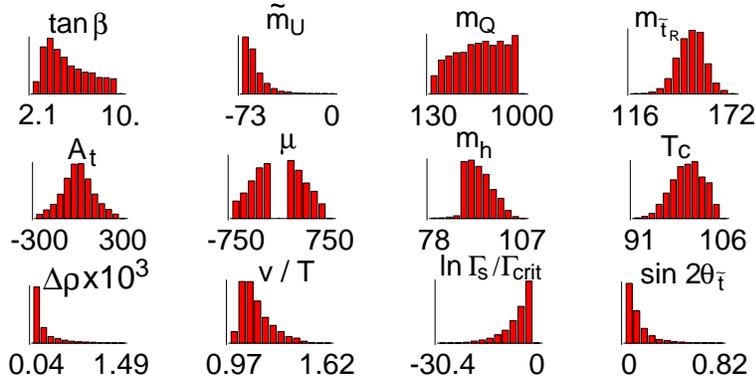,height=2in}}
\caption{Distributions of parameters giving a strong enough electroweak
phase transition for baryogenesis in the MSSM. $\theta_{\tilde t}$ is
the stop mixing angle.  $A_t$ and $\mu$ appear as $y(A_t H_1-\mu H_2)$
in the off-diagonal part of the squark mass matrix.  Units are GeV.}
\label{fig2}
\end{figure}

\begin{figure}[h!] 
\centerline{\epsfig{file=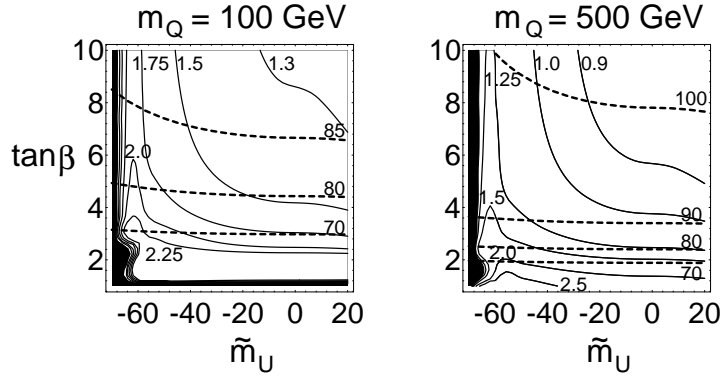,height=2in}}
\caption{Contours of constant $v/T$ (solid lines) and $m_h$ (dashed lines)
in the plane of $\tan\beta$
and $\widetilde m_U$, for two choices of $m_Q$, 100 and 500 GeV.}
\label{fig2.5}
\end{figure}

In figure \ref{fig2} we show the distributions of parameters which pass
all the cuts mentioned above.  The most important of these is the right
stop soft-breaking mass parameter, $m^2_U$.  Since it must be negative to
get a light enough stop and strong enough phase transition, we define
$\widetilde m_U = m^2_U/|m_U|$.  As I will discuss below, too negative a value
of $\widetilde m_U$ gives color breaking vacua, which are excluded.  The dependence 
of the strength of the EWPT on $\widetilde m_U$, as measured by $v/T$,
is shown in figure \ref{fig2.5}, where the tendency for $v/T$ to increase
as $\widetilde m_U$ becomes more negative in clear.  One also sees that
$v/T$ is a decreasing function of $\tan\beta$, but that this can always 
be counteracted by going to more negative $\widetilde m_U$ values.

\vspace{-0.25in}

\subsection{Correlations}

\begin{figure}[h!] 
\centerline{\epsfig{file=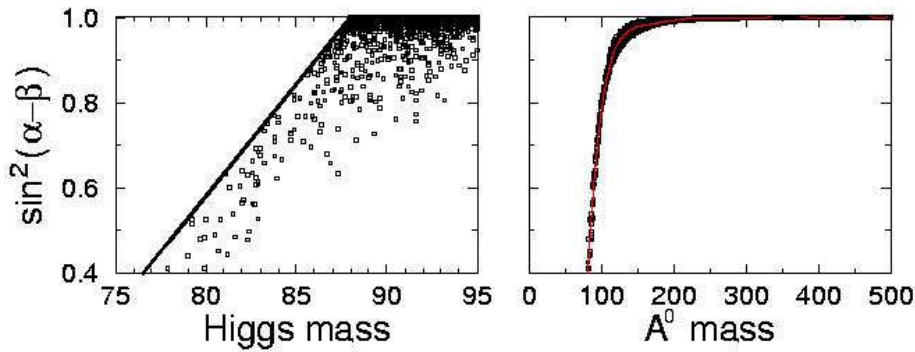,height=2in}}
\caption{Higgs field alignment parameter $\sin^2(\alpha-\beta)$ versus (a) $m_h$ and (b)
$m_{A^0}$, in GeV.}
\label{fig3}
\end{figure}

Perhaps more illuminating are the correlations between parameters.  One
of these, $\sin^2(\alpha-\beta)$, measures the alignment between the
direction of symmetry breaking in the $H_1$-$H_2$ plane and the
direction of the lightest Higgs field.  When $\sin^2(\alpha-\beta) =
1$, one has recovered the limit of a SM-like (single Higgs field) Higgs
sector.  Figure \ref{fig3} shows that most of the accepted points
correspond to this regime, which is also the limit where $m_{A^0}\gg
m_h$.  Recent results from L3 at LEP have increased the experimental
lower limit on the Higgs mass to exclude many of the points shown in
figure \ref{fig3}(a): the limit at $\sin^2(\alpha-\beta)=1$ is now
$m_h>95.5$ GeV. \cite{L3}

\begin{figure}[t!] 
\centerline{\epsfig{file=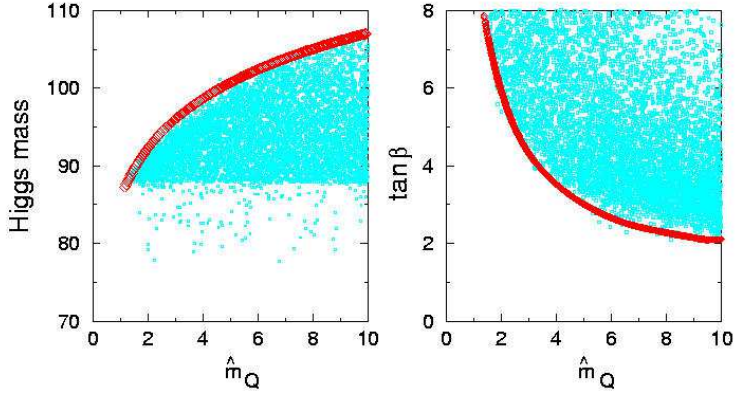,height=2.5in}}
\vspace{-10pt}
\caption{Correlations of (a) $m_h$ and (b) $\tan\beta$ with the
left stop mass, $\hat m_Q = m_Q/(100$ GeV).}
\label{fig4}
\end{figure}

In figure \ref{fig4} we show how $m_h$ and $\tan\beta$ depend upon the
mass of the heavy (left) stop, $m_Q$.  This dependence comes about because
of the radiative corrections to $m_h$, which are roughly proportional to
$y^2 m_t^2 \ln(m_{\tilde t_R} m_Q/ m_t^2)$.  To insure that $m_h$ exceeds
the experimental limit, one can go to large $m_Q$, or alternatively one
can make $\tan\beta$ large, which increases the tree-level contribution to
$m_h$.  The resulting upper limit on $m_h$ or lower limit on $\tan\beta$
can be expressed in terms of $\hat m_Q\equiv m_Q/(100$ GeV) as
\beqa
	m_h &<& 85.8 + 9.2\ln(\hat m_Q) {\rm\ GeV}\nonumber\\
	\tan\beta &>& (0.03 + 0.076\, \hat m_Q - 0.0031\, \hat
	m_Q^2)^{-1}
\eeqa
Although it would seem rather fine-tuned to have $m_Q\gg m_{\tilde t_R}$,
this logical possibility provides a means of pushing $m_h$ just outside
of the experimental limits at the present and even after the end of the
current LEP2 run.

\subsection{Bubble wall profile}

\begin{figure}[h!] 
\centerline{\mbox{\epsfig{file=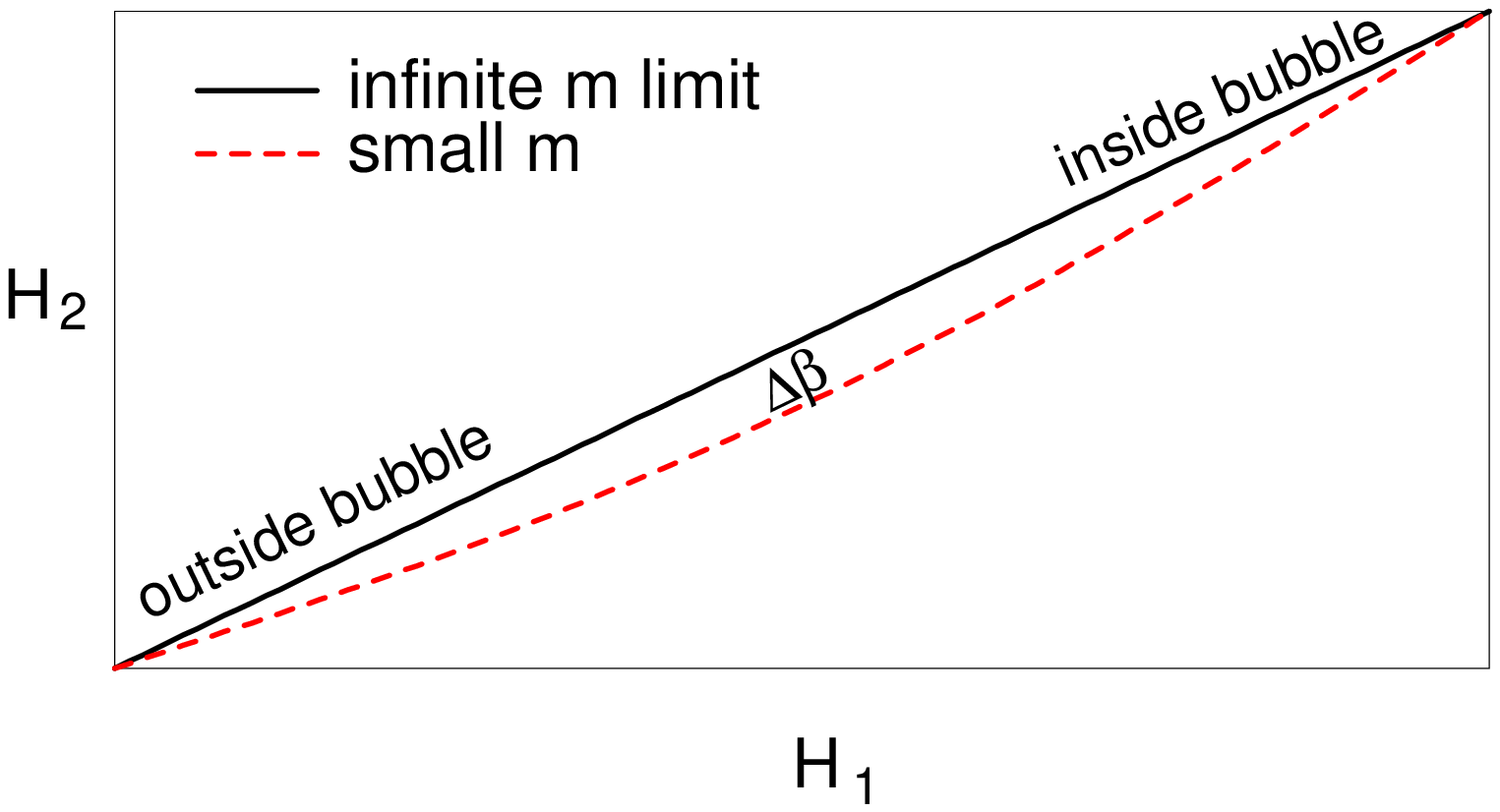,width=3.30in}}
\hskip-0.00in
\mbox{\epsfig{file=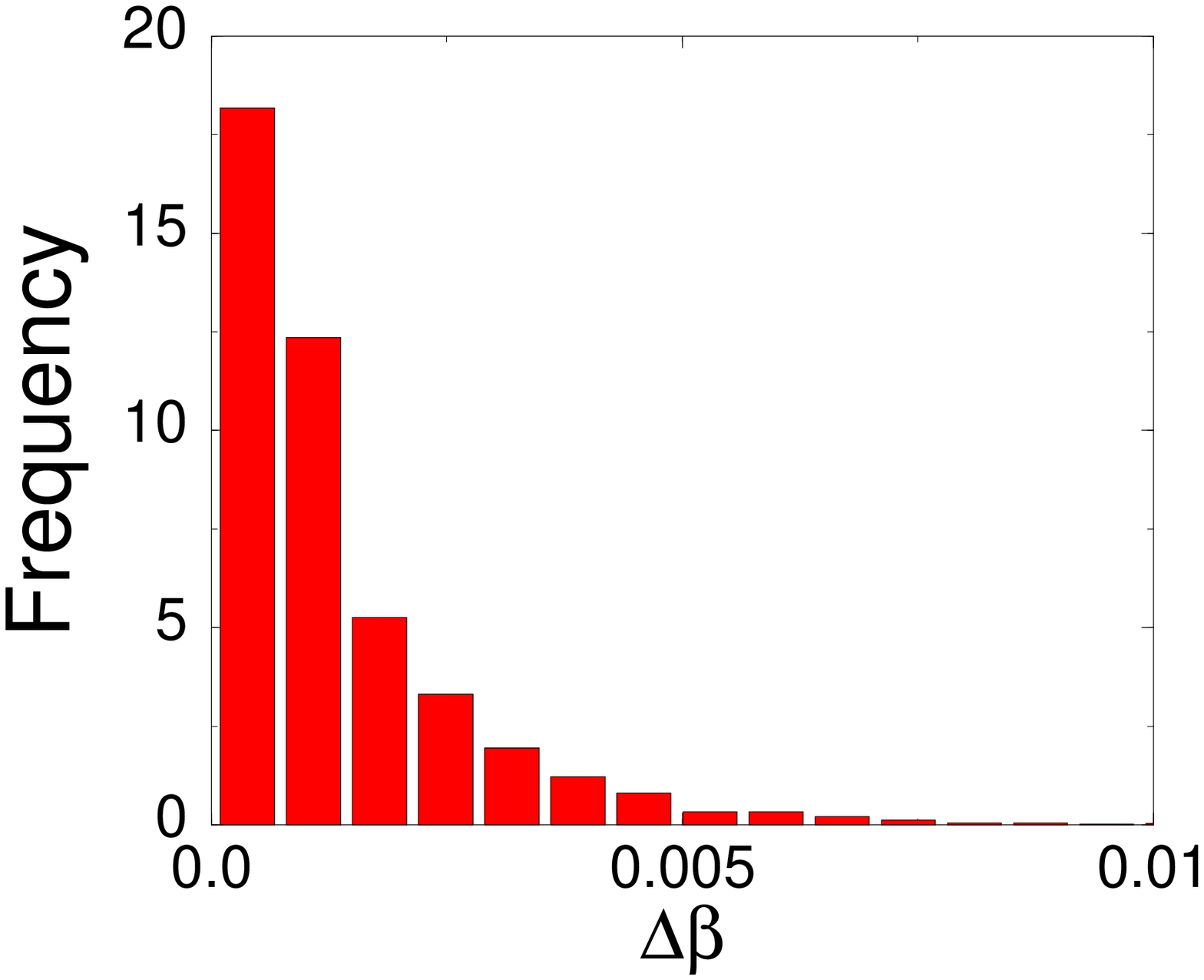,width=2.20in}}}
\vspace{-10pt}
\caption{(a) Higgs field trajectories through the bubble wall
for large and small $m_{A^0}$;\protect\newline (b) Frequency of
$\Delta\beta$ values
from the Monte Carlo of ref.\ [12].}
\label{fig6}
\end{figure}

Another quantity of interest is the deviation of $\tan\beta\equiv
H_2/H_1$ from constancy as a function of distance along the bubble
wall,\cite{MQS} since some contributions to the baryon asymmetry are
suppressed by the change in $\beta$, $\Delta\beta$.\cite{CQRVW,CJK}\ \ 
Figure \ref{fig6}(a) shows that in the large $m_{A^0}$ limit,
$\Delta\beta$ is suppressed, although two-loop effects do enhance
it.\cite{MQS}\ \ Figure \ref{fig6}(b) is the distribution of
$\Delta\beta$ values (which we define somewhat differently than
ref.\ [14]) obtained from our Monte Carlo.\cite{CM}

\section{CP Violation and Transport at the Bubble Wall}

Although the strength of the the EWPT in the MSSM is well understood,
there is less agreement about how to compute the size of the baryon
asymmetry.  The standard approach is to solve a set of diffusion equations
for the density or chemical potential $\mu_L$ of the left-handed
fermions which bias sphalerons.  The baryon asymmetry is related to the
sphaleron rate, the bubble wall velocity, and the integral of this
chemical potential in front of the wall, by
\beq
   n_B \sim {\Gamma_{\rm sph}\over v_w}\int_0^\infty \!\!\mu_L(x)\, dx\,.
\eeq

The problem is how to compute the CP-violating source term in the
diffusion equations for $\mu_L$.  Ref.\ [15] uses a
closed-time-path, nonequilibrium quantum field theory
method;\cite{Riotto} however ref.\ [18] makes some approximations that
have been questioned.\cite{JKP}\ \ In ref.\ [17] we instead
adapted a formalism \cite{JPT} in which the semiclassical
CP-violating force exerted by the wall on the charginos is taken into
account, using the WKB approximation.  This gives a force term in the
Boltzmann equation which is the leading effect in an expansion in
derivatives of the bubble wall.

\begin{figure}[h!] 
\centerline{\epsfig{file=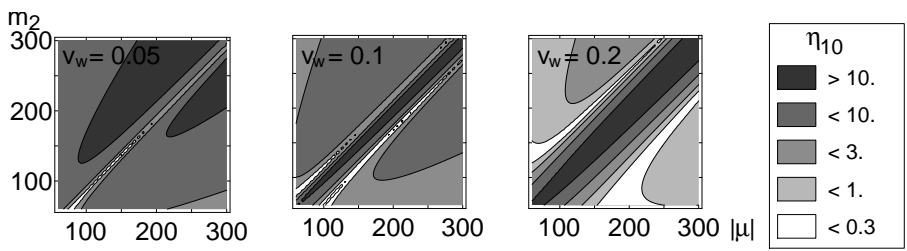,width=4.7in}}
\vspace{-0pt}
\caption{Contours of $\eta_{10}$, the baryon to photon ratio times 
$10^{10}$, in the $m_2$-$\mu$ plane, 
for three different wall velocities, using the classical
force analysis of ref.\ [17].  Masses in GeV.}
\label{fig7}
\end{figure}

Both methods give qualitatively similar results: it is the charginos
which are primarily responsible for the CP asymmetry, because of the CP
violating phase Im$(m_2 \mu)$ of the parameters in the Wino-Higgsino
Dirac mass matrix,
\beq
	{\cal M}_{\tilde W\tilde h} = \left(\begin{array}{cc}
	m_2 & gH_2/\sqrt{2} \\ gH_1/\sqrt{2} & \mu  \end{array}
	\right)\, .
\label{chargino}
\eeq
And they both, roughly speaking, predict a larger baryon asymmetry
when $m_2\cong\mu$.  But quantitatively the two methods disagree; for
example they obtain different dependences on the wall velocity, as
demonstrated in figures \ref{fig7} and 8.

\begin{figure}[t!] 
\centerline{\mbox{\epsfig{file=cqw.epsi,width=2.0in}}
\hskip0.20in
\mbox{\begin{minipage}{2.5truein}\vspace{-2.4in}
{\small
Figure 8: Contours of constant CP-violating phase needed
to get the observed baryon asymmetry, in the $m_2$-$\mu$ plane, 
from ref.\ [15].  The region in the center of the ellipse
requires the smallest
value of the phase, $\phi =  0.06$, and thus represents the most
efficient parameters for generating the baryon asymmetry.}
\end{minipage}}}
\label{fig8}
\end{figure}

A second discrepancy between references [15] and [17] is that the WKB
analysis shows no suppression in the result proportional to
$\Delta\beta$ (discussed in the previous section), whereas ref.\ [15]'s
result is proportional to $\Delta\beta$.  A third difference is that we
do find that $\mu_L$ is suppressed by helicity-flipping interactions of
Higgsinos in front of the wall, involving the $\mu$ term of
eq.~(\ref{chargino}), while ref.\ [15] does not.  For the moment these
disagreements remain unresolved.

\section{Was SU(3)$_{\rm color}$ broken just before the EWPT?}

We have seen that the right stop must be very light for electroweak
baryogenesis to work in the MSSM.  This leads to the interesting
possibility that the SU(3) gauge group of QCD was temporarily broken
before the EWPT.\cite{KLS,BJLS,CQW2}\ \ Because the right-stop
soft-breaking mass $m^2_U$ is negative in eq.\ (\ref{stopmass}), the
symmetric vacuum is unstable toward condensation of the stop field in
some random direction in color space.  If $\widetilde m_U$ is
sufficiently negative, it can be energetically preferable to have a
period of color breaking before the normal electroweak vacuum state
takes over.  Then one could have the sequence
\def\aone{\mathrel{\hbox{\rlap{\hbox{\lower10pt\hbox{$\,1$}}}\hbox{$\to$}}}}
\def\atwo{\mathrel{\hbox{\rlap{\hbox{\lower10pt\hbox{$\,2$}}}\hbox{$\to$}}}}
\beq
	(H,\ \tilde t_R) \ = \ (0,\ 0) {\,\aone\,} (0,\, v_{\tilde t})
 {\,\atwo\,}
	(v_h,\, 0) \eeq of phase transitions, which would change our
view of cosmological history in an interesting way.  For example, the
second transition tends to be very strong, which is favorable for
baryogenesis.  But there is an energy barrier, shown in figure
9(a), which impedes this second stage of the transition, due to a
term in the potential \beq
	y^2 |H_2|^2 |\tilde t_R|^2 
\eeq 
whose positivity (provided squark
mixing is not too strong) is a consequence of supersymmetry.  It could
happen that the rate of tunneling from the color-broken to the
electroweak phase is so small that it will never happen in the history
of the universe.  Guy Moore\footnote{and ref.\ [21], as was brought
to our attention} has made the following conjecture: if $m_{\tilde
t_r}$ is ever small enough for transition 1 to take place, then the
universe gets stuck in the color broken phase and never completes
transition 2.  Although preliminary studies of this question have been
done,\cite{BJLS,CQW2} it deserves a more careful treatment.

\addtocounter{figure}{1}
\begin{figure}[b!] 
\centerline{\mbox{\epsfig{file=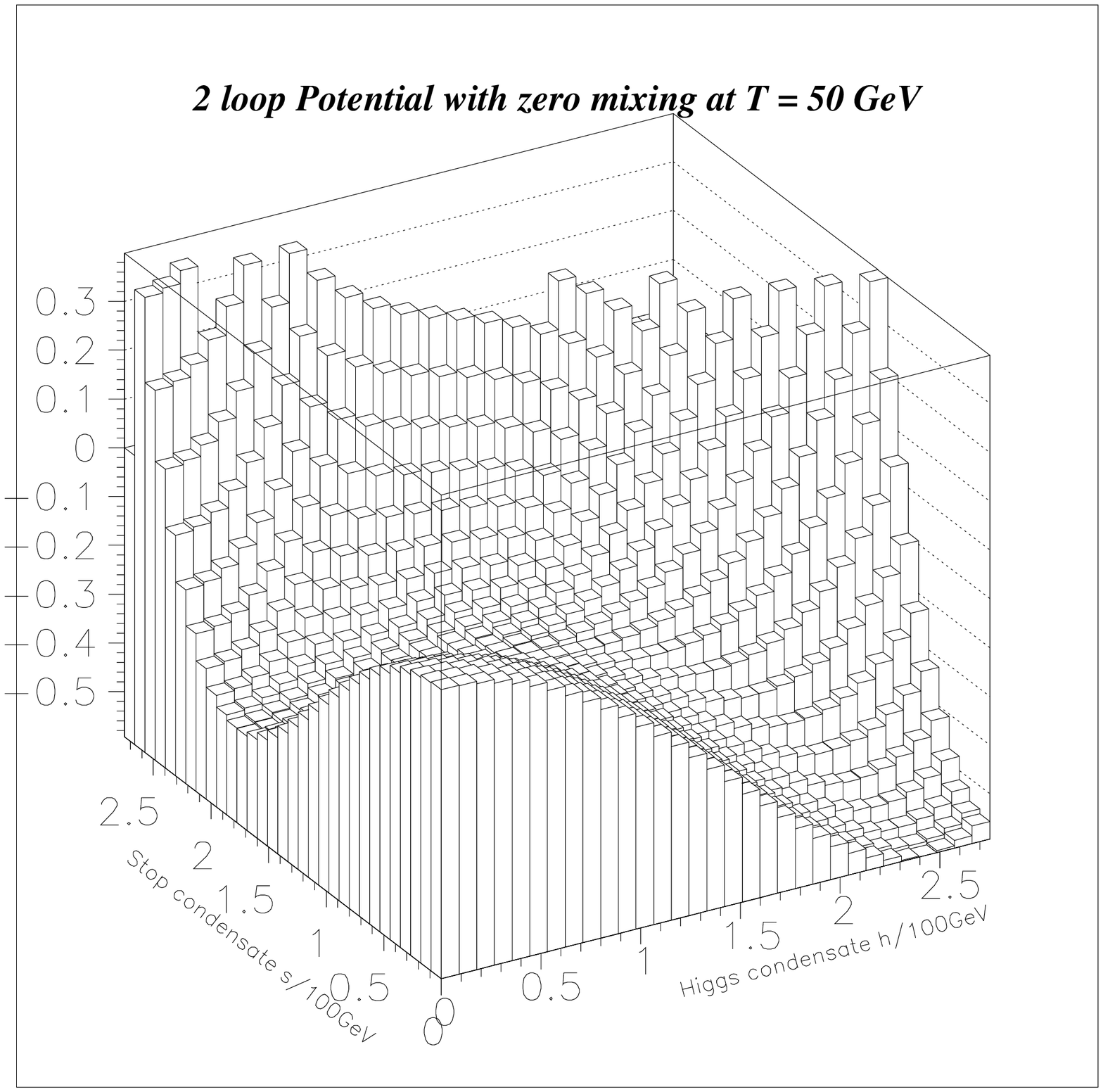,width=2.0in}}
\hskip0.00in
\mbox{\epsfig{file=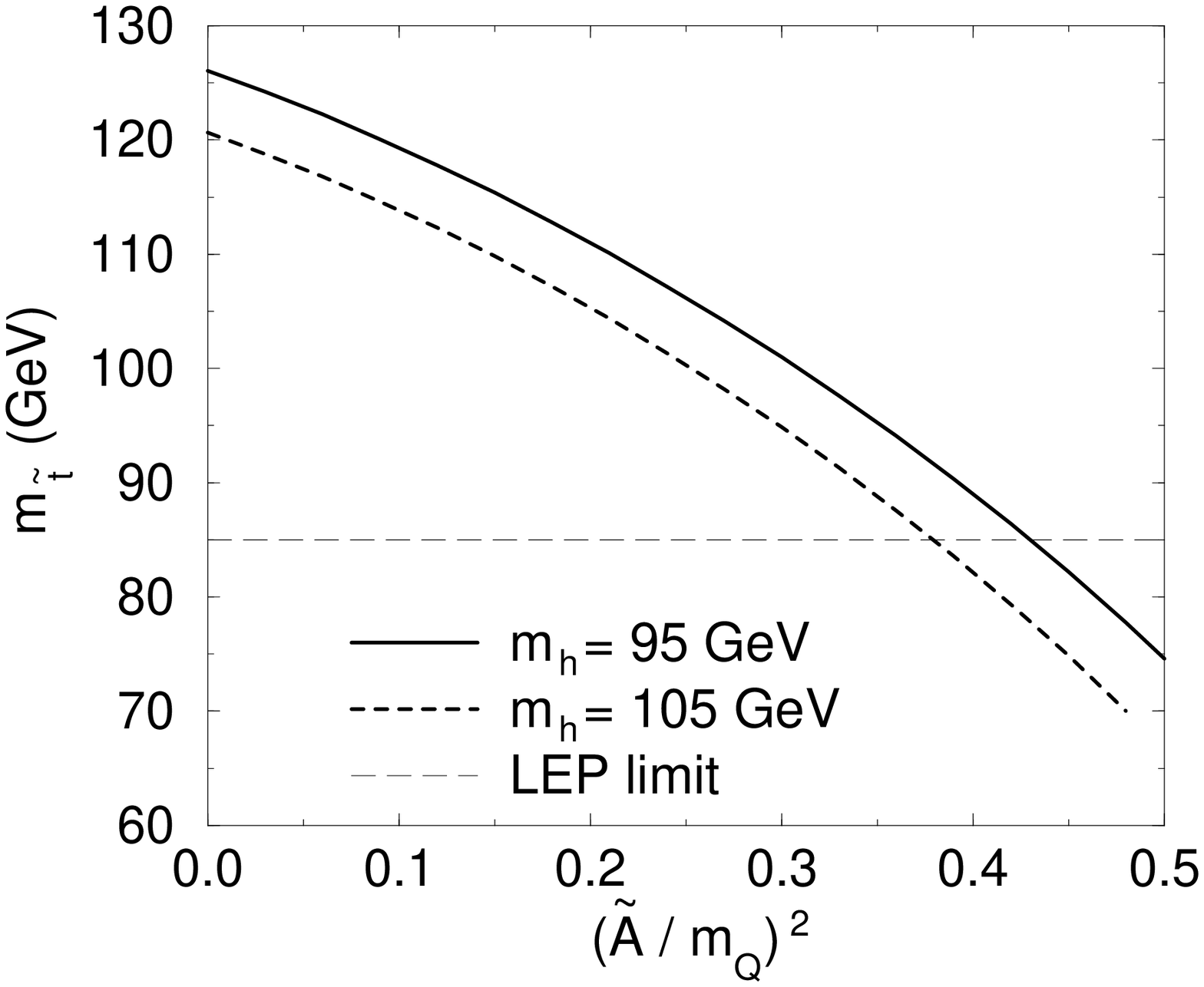,width=2.5in,height = 2.1in}}
}
\label{fig9}
\vspace{-0pt}
\caption{(a) The two-loop effective potential $V(H,\tilde t_R)$ for the
Higgs field (right axis) and right stop (left axis), at the temperature
$T = 50$ GeV where the tunneling rate from the metastable CCB minimum
to the electroweak vacuum is maximized.  (b) Lower limit on $m_{\tilde
t_R}$ to avoid color breaking, as a function of {\scriptsize $(A_t -
\mu\cot\beta)^2/m_Q^2$}, for two values of $m_h$.  } 
\end{figure}

We have undertaken such a study, by constructing the two-loop
effective potential $V(H,\,\tilde t_R)$ for the Higgs and stop fields,
and computing the nucleation rate for the most likely bubbles
interpolating between the color-broken and electroweak
phases.\cite{CMS}\ \ This involves finding the path in the $(H,\,\tilde
t_R)$ field space along which the bubble evolves, which gives the
lowest bubble energy, hence the fastest rate of transitions.  The field
equations with boundary conditions $(0,\,v_{\tilde t})$ and $(v_h,\,0)$
at the respective ends must be solved along this path.  One needs a
value of bubble energy over temperature smaller than $E/T \cong 170$ to
get a tunneling rate per unit volume, $\Gamma\sim T^4 e^{-E/T}$, that
is competitive with the Hubble rate per Hubble volume, ${\cal H}^4$.
That is, $E/T$ must be less than $4 \ln(T/{\cal H})$.  However, we find
that even when all MSSM parameters are adjusted to the values that are
optimal for tunneling, the exponent $E/T$ is too large by a factor of
5.  This allows one to rule out color breaking at the EWPT.  Any
resulting constraints must be made in a high-dimensional space of
parameters.  As an example we show the lower limit on the right stop
mass as a function of the squark mixing parameter $\tilde A/m_Q$ in
figure 9(b).

It therefore appears that one must go beyond the MSSM in order to make
color-breaking a real possibility just before the electroweak phase
transition.  However it may be possible to change this conclusion in
extended models, such as those with $R$-parity breaking, if one can 
arrange for large negative corrections to the Debye mass of the light
stop.\cite{CMS}

\end{document}